\documentclass[aps,prb,twocolumn,superscriptaddress]{revtex4}

\usepackage{graphicx,color}

\input epsf.sty
\usepackage{graphicx}

\begin{document}

\preprint{\today}


\title{
Magnetic field effect on Fe-induced short-range magnetic correlation and electrical conductivity in Bi$_{1.75}$Pb$_{0.35}$Sr$_{1.90}$Cu$_{0.91}$Fe$_{0.09}$O$_{6+y}$
}

\author{Shuichi Wakimoto\footnote{Corresponding author: wakimoto.shuichi@jaea.go.jp}}
\affiliation{ Quantum Beam Science Directorate, Japan Atomic Energy Agency,
   Tokai, Ibaraki 319-1195, Japan }

\author{Haruhiro Hiraka}
\affiliation{ Institute for Materials Research, Tohoku University, Katahira,
   Sendai 980-8577, Japan }

\author{Kazutaka Kudo\footnote{Present address: Department of Physics, Faculty of Science, Okayama University, Okayama, Japan.}}
\affiliation{ Institute for Materials Research, Tohoku University, Katahira,
   Sendai 980-8577, Japan }

\author{Daichi Okamoto}
\affiliation{ Institute for Materials Research, Tohoku University, Katahira,
   Sendai 980-8577, Japan }

\author{Terukazu Nishizaki}
\affiliation{ Institute for Materials Research, Tohoku University, Katahira,
   Sendai 980-8577, Japan }

\author{Kazuhisa Kakurai}
\affiliation{ Quantum Beam Science Directorate, Japan Atomic Energy Agency,
   Tokai, Ibaraki 319-1195, Japan }

\author{Tao Hong}
\affiliation{ Neutron Scattering Science Division, Oak Ridge National
   Laboratory, Oak Ridge, Tennessee 37831, USA}

\author{Andrey Zheludev\footnote{Present address: Laboratorium f\"ur Festk\"orperphysik, ETH H\"onggerberg 8093 Z\"urich, Switzerland.}}
\affiliation{ Neutron Scattering Science Division, Oak Ridge National
   Laboratory, Oak Ridge, Tennessee 37831, USA}

\author{John M. Tranquada}
\affiliation{ Condensed Matter Physics \& Materials Science Department, 
   Brookhaven National Laboratory, Upton, New York 11973-5000, USA  }

\author{Norio Kobayashi}
\affiliation{ Institute for Materials Research, Tohoku University, Katahira,
   Sendai 980-8577, Japan }

\author{Kazuyoshi Yamada}
\affiliation{ Institute for Materials Research, Tohoku University, Katahira,
   Sendai 980-8577, Japan }
\affiliation{ WPI Research Center, Advanced Institute for Materials Research, 
   Katahira, Sendai 980-8577, Japan}

\date{\today}

\begin{abstract}

We report electrical resistivity measurements and neutron diffraction studies under magnetic fields of 
Bi$_{1.75}$Pb$_{0.35}$Sr$_{1.90}$Cu$_{0.91}$Fe$_{0.09}$O$_{6+y}$,
in which hole carriers are overdoped.
This compound shows short-range incommensurate magnetic correlation with incommensurability $\delta=0.21$, whereas a Fe-free compound shows no magnetic correlation.  
Resistivity shows an up turn at low temperature in the form of $\ln(1/T)$ and shows no superconductivity.
We observe reduction of resistivity by applying magnetic fields ({\it i.e.}, a negative magnetoresistive effect) at temperatures below the onset of short-range magnetic correlation.   Application of magnetic fields also suppresses the Fe-induced incommensurate magnetic correlation.   We compare and contrast these observations with two different models: 1) stripe order, and 2) dilute magnetic moments in a metallic alloy, with associated Kondo behavior.  The latter picture appears to be more relevant to the present results.

\end{abstract}

\pacs{74.72.Gh, 74.25F-, 75.47.Np, 75.25.-j}

\maketitle

\section{Introduction}

Incommensurately-modulated antiferromagnetic (AF) correlations appear to be a universal feature in the hole-doped high-temperature superconducting cuprates La$_{2-x}$Sr$_{x}$CuO$_{4}$ (LSCO) and YBa$_{2}$Cu$_{3}$O$_{6+y}$.~\cite{Birgeneau_JPSJ06} 
Neutron-scattering measurements of the low-energy dynamical AF correlations in underdoped samples reveal peaks at the wave vectors $(0.5\pm\delta, 0.5)$ and $(0.5, 0.5\pm\delta)$ in reciprocal lattice units (r.l.u.) of the CuO$_{2}$ square lattice (1 r.l.u. $\sim 1.67$~\AA$~{-1}$).  Here $\delta$ is referred to as an incommensurability.  These magnetic signals disperse inwards, become commensurate at a intermediate energy range, and then disperse outwards, exhibiting the characteristic ``hour-glass'' shaped spectrum.~\cite{hayd04,tran04}
In LSCO, such low-energy incommensurate (IC) magnetic correlations disappear coincidently with the disappearance of the superconductivity in the heavily overdoped region, indicating a strong correlation between superconductivity and AF correlations.~\cite{waki_04,waki_07,lips_07}  However, in cuprates with higher $T_c$'s, such as Bi-, Tl-, and Hg-systems, reports of magnetic fluctuations are very sparse~\cite{xu_cm09} and clear IC AF correlations have not been observed, whereas the commensurate resonant magnetic scattering in the superconducting state has been observed.~\cite{Bi,Tl}  Thus, there are gaps in the experimental record relevant to understanding the role of magnetism in the mechanism of superconductivity.

Very recently, it has been reported that the replacement of a small amount of Cu sites by magnetic Fe$^{3+}$ ions in the La$_{2-x}$Sr$_{x}$CuO$_{4}$ (LSCO) and overdoped Bi$_{1.75}$Pb$_{0.35}$Sr$_{1.90}$CuO$_{6+y}$ (BPSCO) systems dramatically stabilizes a short-range (SR) IC magnetic correlation.~\cite{Fujita_09CM,hira_09}  In the latter case, neutron-scattering measurements clarified the SR IC magnetic correlation below 40 K,~\cite{hira_09} unveiling potential IC spin correlations in the single-layered Bi-based cuprate system.  Remarkably, the observed incommensurability is $\delta = 0.21(1)$~r.l.u.  This value is approximately equal to the effective hole concentration of the sample, consistent with a trend found in underdoped LSCO, but greatly exceeding the well-known saturation limit of $\sim0.125$ r.l.u. in the LSCO system.~\cite{Yamada_98}

There have been a number of reports that non-magnetic and magnetic impurities, such as Zn and Ni, substituted for Cu atoms in superconducting LSCO stabilize IC magnetic order.~\cite{zn1,zn2,ni1,ni2} 
The induced order is attributed to a static stripe order resulting from trapping of hole carriers by the impurities.  
On the other hand, the disappearance of low-energy IC magnetic correlations in the overdoped LSCO~\cite{waki_07} implies no stripe correlation to be stabilized by impurities in the overdoped region.  Thus, the newly found IC magnetic correlation in Fe-doped overdoped BPSCO may have a different origin from stripe order.
In order to clarify this issue, we performed resistivity and neutron diffraction studies under magnetic fields, since the effect of magnetic fields on the stripe magnetic order has been well studied in LSCO and related compounds.
This problem should give new insights into the origin of the IC magnetic correlations of the high-temperature superconducting cuprates in the overdoped metallic regime.

We find that the sample shows a negative magnetoresistive effect: that is, the resistivity decreases when a magnetic field is applied along the $c$-axis.  This effect grows below 40~K where the SR IC magnetic correlation sets in.  Neutron diffraction reveals that an applied magnetic field slightly reduces the IC magnetic correlation.  These effects are in contrast with the effect of magnetic fields on stripes in the LSCO system, in which the stripes with $\delta\sim0.12$~r.l.u. are generally stable in magnetic fields; in fact, static order is typically enhanced by an applied field in the underdoped regime.\cite{Katano_00,Lake_02,Boris_02,Chang_08} 
As an alternative, dilute magnetic moments in a metallic alloy may be a relevant model for the present case.  Fe spins in the metallic background in the overdoped BPSCO may behave as Kondo scatterers.  This induces the Kondo effect in the resistivity and magnetic correlation appears by the RKKY interaction.  
Magnetic fields compete with the exchange coupling between conduction electrons and impurity moments in Kondo systems, and such an effect might explain our observation that the Fe-induced magnetic order is reduced by magnetic fields.

\section{Experimental details}

A single-crystal of Bi$_{1.75}$Pb$_{0.35}$Sr$_{1.90}$Cu$_{0.91}$Fe$_{0.09}$O$_{6+y}$ (Fe-doped BPSCO) used in the neutron scattering experiments is identical to that studied in Ref.~\onlinecite{hira_09}.  Hole concentration estimated from Fermi surface area measured by angle-resolved photoemission spectroscopy (ARPES)~\cite{Sato_unpub} is $p \sim 0.23$; thus, the hole carriers are overdoped in the present sample.  Small pieces cut from the same crystal rod were used for the magnetization and resistivity measurements.  The sample shows no superconductivity down to 1.6~K.  From the neutron scattering measurements, the crystal structure is orthorhombic down to the lowest temperature that we measured.  Lattice constants are $a=5.30$~\AA~ and $b=5.37$~\AA~ at room temperature, with a corresponding lattice unit of the CuO$_2$ square lattice of $3.77$~\AA.

Magnetization measurements were performed using a superconducting quantum interference device (SQUID) magnetometer.  A crystal with typical size of $2 \times 2 \times 2$~mm$^3$ was fixed using a plastic straw.  
For the in-plane resistivity measurements, single crystals were cut and shaped with typical size of $2.0 \times 0.8 \times 0.03$~mm$^3$.  Then, four electrodes were attached by heating hand-painted gold-paste at $300^\circ$C, and samples were wired by silver paste and gold wires.  After these procedures, the in-plane resistivity measurements under magnetic fields were carried out by a standard DC four-terminal method.  Magnetic fields were applied up to 15 T parallel to the $c$-axis by a superconducting magnet. 

Neutron diffraction measurements were carried out using the TAS-1 triple-axis spectrometer at the research reactor JRR-3, Tokai, Japan, and the HB-1 triple-axis spectrometer at the High Flux Isotope Reactor (HFIR), Oak Ridge National Laboratory, USA.
For both instruments, the wavelength, $\lambda$, of incident neutrons was selected by a pyrolytic graphite (PG) monochromator, while the energy of diffracted neutrons was determined by a PG analyzer.  A PG filter was placed in the neutron path to eliminate neutrons with wavelengths of $\lambda/2, \lambda/3$, and so on. 
The TAS-1 spectrometer was used for measurements without magnetic fields, using neutrons with $\lambda=2.359$~\AA, corresponding to the energy $E = 14.7$~meV.  Horizontal divergences of neutron beam upstream and downstream of monochromator and similarly for the analyzer were $40'$, $80'$, $80'$, and $240'$, respectively.
The HB-1 spectrometer was used for the measurements with magnetic fields up to 5~T, applied parallel to the $c$-axis of the sample, using neutrons with $\lambda=2.462$~\AA, corresponding to the energy $E = 13.5$~meV.  Horizontal beam divergences are $40'$, $60'$, $80'$, and $200'$.
The Fe-doped BPSCO crystal was mounted in a refrigerator with the $a$ and $b$ axes laid in the horizontal neutron scattering plane, so that the $(H, K, 0)$ reflections were accessible.  In the present paper, we will use orthorhombic notation, in which the antiferromagnetic propagation vector of the basal CuO$_2$ plane corresponds to either $(1, 0, 0)$ or $(0, 1, 0)$.

\section{Magnetization}

\begin{figure}
\includegraphics[width=8cm]{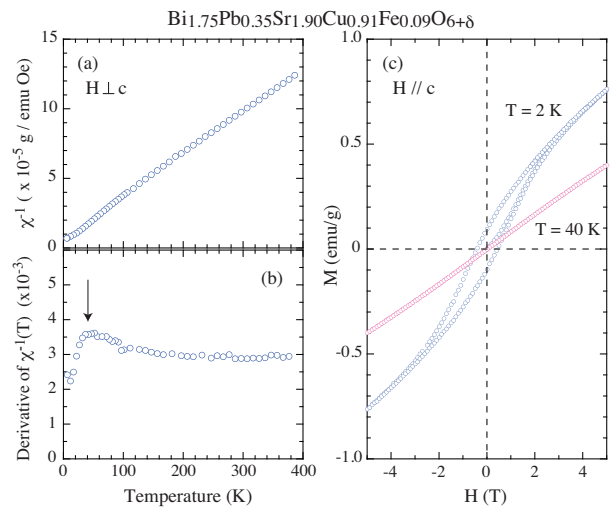}
\caption{(Color online) Temperature dependence of (a) inverse of in-plane magnetic susceptibility, $\chi^{-1}$ and (b) its derivative of Fe 9\%-doped Bi2201.  The arrow shows $T_m$ that is magnetic ordering temperature determined by neutron scattering.  (c) $M-H$ curve at 2~K and 40~K measured after cooling in zero field.}
\end{figure}

We first present the results of the magnetization measurements.  
Figure 1(a) shows temperature ($T$) dependence of the inverse magnetic susceptibility $\chi^{-1}$ measured in a 1~T magnetic field applied perpendicular to the $c$-axis.  Here, $\chi$ has been evaluated by subtracting the background magnetic susceptibility arising from the sample holder (plastic straw).  In the high temperature region, $\chi^{-1}$ is nearly linear in $T$; therefore, the system shows paramagnetism following the Curie-Weiss law.  The extrapolated negative intercept of the temperature axis suggests weakly antiferromagnetic spin correlations.
This Curie-Weiss behavior comes dominantly from doped Fe spins, as the previous study~\cite{hira_09} indicates that the susceptibility increases linearly with the Fe concentration.
However, at low temperatures, $\chi^{-1}$ deviates from the Curie law.  This is clearly seen in Fig. 1(b), where the derivative of $\chi^{-1}$ is plotted.  As $T$ decreases, the derivative reaches a maximum at $\sim 40$~K, and starts to decrease below that.
On the other hand, this compound shows true spin freezing at $T_{sg} \sim 9$~K into a cluster spin glass state, in which spins in the each cluster are antiferromagnetically correlated.  The deviation from the Curie law below $\sim 40$~K might be due to a precursor of the cluster glass state, since the temperature of 40~K is close to the point where SR magnetic correlation sets in, as we will show later.

At very low temperatures, $T < T_{\rm sg}$, the system shows a weak ferromagnetic character.
Figure 1(c) shows $M$--$H$ curves at 2 and 40 K measured with $H // c$ after cooling in zero field.  The data at 40 K demonstrate that the system is paramagnetic, whereas the data at 2 K exhibit a small hysteresis loop, indicating that the system has a weak ferromagnetic response along the $c$-axis. %
We confirmed that there is no hysteresis in the in-plane magnetization at 2~K.  We will return to the nature of the magnetic interactions and the relevance of these results in Sec.~\ref{sc:disc}.

\section{Resistivity}

\begin{figure}
\includegraphics[width=8cm]{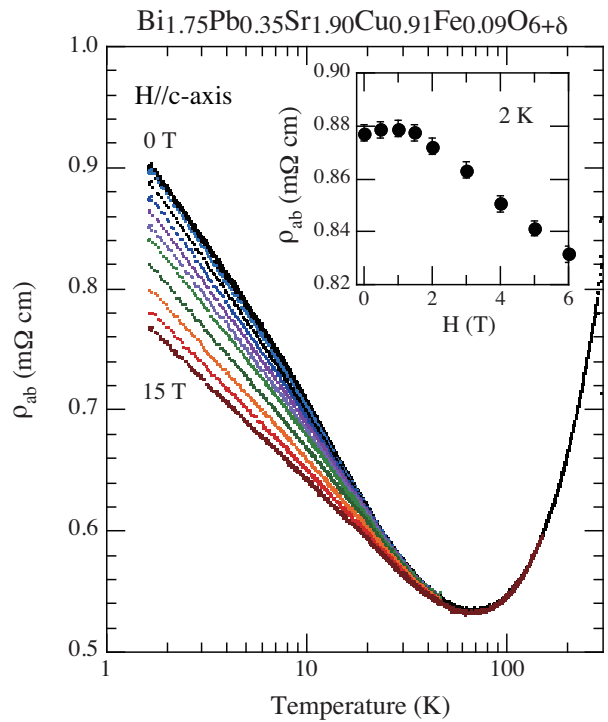}
\caption{(Color online) Temperature dependence of the in-plane resistivity $\rho_{ab}$ under various magnetic fields at 0, 0.5, 1, 1.5, 2, 3, 4, 5, 6, 7, 9, 11, 13, and 15~T along the $c$-axis.  Inset figure shows $H$-dependence of $\rho_{ab}$ at 2~K.}
\end{figure}

Figure 2 shows the $T$-dependence of the in-plane resistivity $\rho_{ab}$ measured in various magnetic fields.  The system shows an up turn at low temperature in the form of $\ln(1/T)$, and shows no superconductivity down to the lowest temperature, 1.6~K.  For $0 \leq H \leq 2$~T, the data sets overlap, while for $H > 2$~T and $T \alt 40$~K, $\rho_{ab}$ is reduced by magnetic fields, thus demonstrating a negative magnetoresistive effect.  The inset shows $\rho_{ab}$ as a function of magnetic field at the lowest temperature.  It shows that the negative magnetoresistive effect appears for $H \agt 2$~T.

\begin{figure}
\includegraphics[width=8cm]{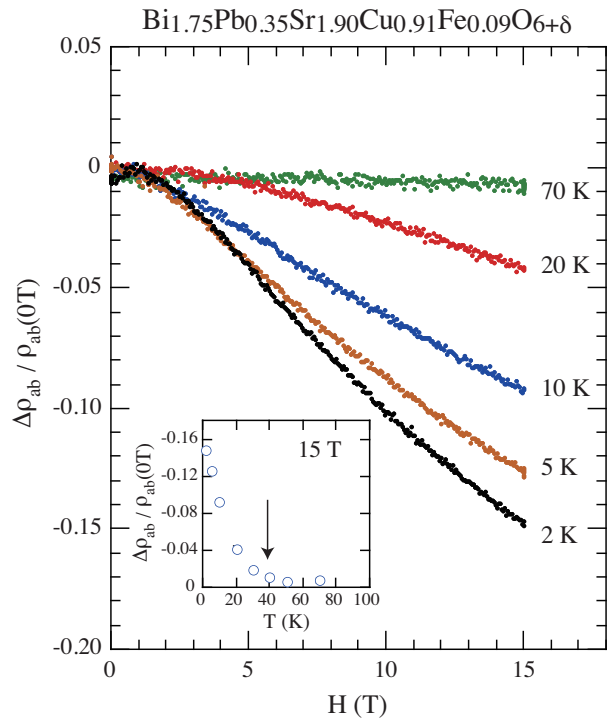}
\caption{(Color online) Magnetic field dependence of $\Delta\rho_{ab}/\rho_{ab}(H=0T)$ at selected temperatures. $\Delta\rho_{ab}$ is the reduction of $\rho_{ab}$ by magnetic fields.  The inset shows $\Delta\rho_{ab}(H=15T)/\rho_{ab}(H=0T)$ as a function of temperature.}
\end{figure}

The negative magnetoresistive effect is seen only at low temperatures.  In Fig.~3, we plot $\Delta\rho_{ab}/\rho_{ab}(H=0{\rm T})$ as a function of $H$, where $\Delta\rho_{ab}$ is the change in resistivity caused by magnetic fields.  At 70~K, $\Delta\rho_{ab}$ is nearly zero up to 15~T, and $|\Delta\rho_{ab}|$ increases as temperature decreases.  The inset shows the $T$-dependence of $\Delta\rho_{ab}/\rho_{ab}(H=0{\rm T})$ at 15~T.  It is shown that the negative magnetoresistive effect becomes prominent below 40~K, which is, again, close to the temperature where the SR magnetic correlation sets in.  This fact indicates that the SR magnetic correlation is closely related to the negative magnetoresistive effect.

\section{Neutron diffraction}

Results of the neutron diffraction study in zero field, performed at the TAS-1 spectrometer, are summarized in Fig.~4.  
Figure 4 (a) shows neutron diffraction intensity measured at reciprocal lattice points of $(H, K, L)=(1+q, -q, 0)$ in the orthorhombic notation.  
The difference between the data at 4~K and 80~K is plotted in Fig.~4(b).  
The solid line is a result of fit by resolution-convoluted two-dimensional Lorentzian.  The fitting gives peaks at $q = 0.205(5)$.
Note that the position of $(1, 0, 0)$ and the deviation defined as $(q, -q, 0)$ are equivalent to the antiferrromagnetic wave vector $(0.5, 0.5, 0)$ and the deviation of $(q, 0, 0)$ in reciprocal lattice units of the CuO$_2$ square lattice, as illustrated in Fig.~4(e).  Thus, the incommensurability $\delta$ is 0.205(5).

In identifying the magnetic scattering, we have focused on temperature-dependent features in order to discriminate from a number of spurious peaks that are present in the raw data.  (The same approach was used in Ref.~\onlinecite{hira_09}.)
The sharp peak at $q=0$ in Fig.~4(a) may be due to either multiple scattering or the contribution of a small fraction of neutrons with $\lambda/2$ diffracted from $(2, 0, 0)$.  Other sharp peaks at $q=-0.35$ and $0.5$ may originate from nuclear peaks of small, misoriented grains, as the sample is a mosaic crystal with imperfections; note that the intensities of all peaks in Fig.~4(a) are comparable to the background level.  

We note that the background of the difference plot, Fig.~4(b), has a finite value, suggesting a $T$-dependent background scattering.  To examine this effect, we compare the neutron intensities at $q=-0.2$ and $-0.45$, where the former is located at the top of one IC peak and the latter is far-off from that peak.  Figure 4 (c) exhibits the $T$-dependence of the intensities at these positions.  Even at $q=-0.45$ the intensity increases moderately with decreasing temperature.  This might be attributed to a paramagnetic scattering from Fe spins, which increases with decreasing $T$.  As we saw in Fig.~1, the system indeed shows paramagnetic behavior in its magnetization.  Thus, the $T$-dependence of the IC magnetic peak should be given by subtracting this paramagnetic component.  Figure 4 (d) indicates the difference between intensities at $q=-0.2$ and $-0.45$.  The SR IC magnetic correlation clearly sets in at 40~K, which we define as $T_m$.  
We note that there are small peak structures at $q \sim \pm 0.2$ even at 80~K in Fig. 4 (a).  However, the $T$-dependence in Fig.~4(d) demonstrates that these peaks are independent of $T$ up to 200~K, suggesting that their origin is not magnetic.

The effect of magnetic field on the IC magnetic peak is shown in Fig. 5, which shows the difference in neutron intensity at 5~K and 70~K.  The zero-field data are shown by open symbols and the data with a magnetic field of 5~T are indicated by closed symbols.  For the measurement under magnetic field, the magnetic field of 5~T was applied at 5~K after cooling in zero-field.  Then data were collected at 5~K and 70~K; the sample was heated with the magnetic field kept at 5~T.  
The solid lines are results of fits to resolution-convoluted two-dimensional Lorentzians with background fixed at zero.  Fits give $\delta=0.221(16)$ for the data at $0$~T, and $\delta=0.247(35)$ for the data at $5$~T.
The results show that the IC magnetic peaks have no tendency to be enhanced by the application of magnetic fields.  On the contrary, it appears that magnetic peaks decrease slightly by magnetic fields.

\begin{figure}
\includegraphics[width=8cm]{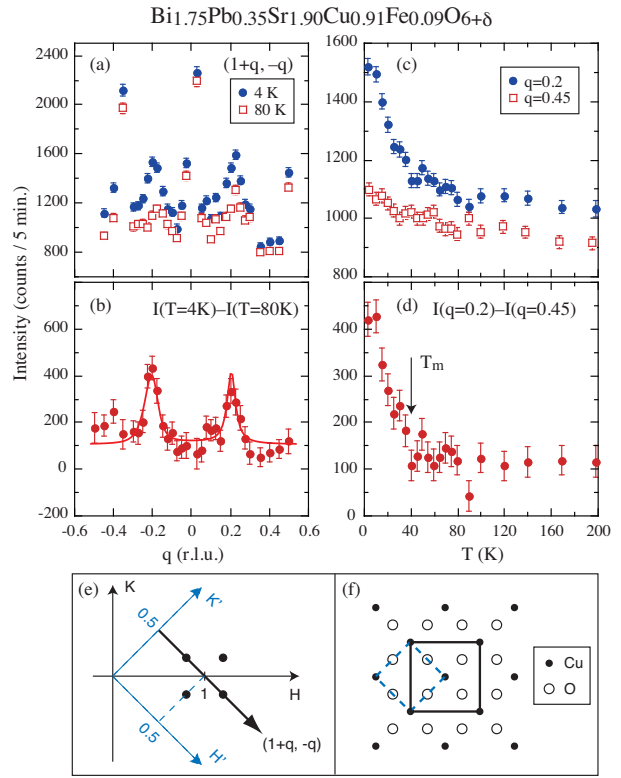}
\caption{(Color online) (a) Neutron scattering intensity on the trajectory of $(1+q, -q)$ at 4 and 80~K without magnetic field measured at the TAS-1 spectrometer.  (b) Difference of neutron intensities at 4~K and 80~K. 
The solid line is a result of fit to a Lorentzian function convoluted with the instrumental resolution.
(c) Temperature dependence of neutron scattering intensity at $q=0.2$ (peak position) and $q=0.45$ (off-peak position).  (d) Difference between intensities at $q=0.2$ and $0.45$ as a function of temperature.  
(e) Incommensurate peak geometry in the reciprocal lattice units.  Four spots around $(H, K)=(1, 0)$ represent IC magnetic peaks.  The arrow indicates the scan trajectory of $(1+q, -q)$.  The axes labeled as $H$ and $K$ define the reciprocal lattice in the orthorhombic notation.  Those labeled as $H'$ and $K'$ define the reciprocal lattice of the CuO$_2$ square lattice.  (f) CuO$_2$ square lattice.  The square by solid line is the orthorhombic unit cell.  The square by dashed line is the unit of the CuO$_2$ square lattice. 
}
\end{figure}

\section{Discussion}
\label{sc:disc}

In this section, we discuss the possible origin of the SR IC magnetic correlations induced by the Fe-doping in the overdoped BPSCO system.

\begin{figure}
\includegraphics[width=8cm]{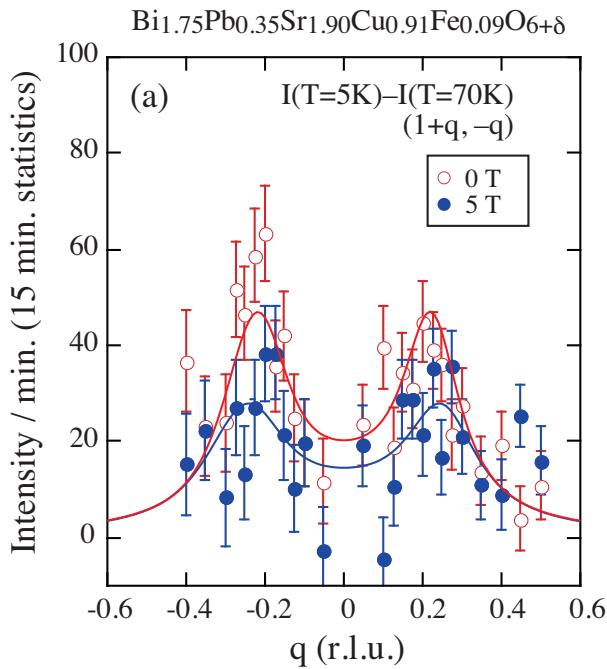}
\caption{(Color online) Difference of neutron scattering intensities at 5~K and 70~K measured at the HB-1 spectrometer under magnetic fields of 0~T and 5~T along the $c$-axis.
Solid lines are results of fits to a Lorentzian function convoluted with instrumental resolution.
}
\end{figure}

\subsection{RKKY interaction}

Our results show that the SR IC magnetic correlation induced by the Fe-doping in the overdoped BPSCO system tends to be slightly suppressed by an external magnetic field. In turn the system gains electron conductivity by an external magnetic field, showing a negative magnetoresistive effect.  In addition, the spin system freezes into a glass state at low temperature.

A model to consider is that of dilute magnetic alloys, such as Cu alloyed with a few percent of Mn.  Beyond the direct analogy of dilute magnetic moments doped into a metallic system, these systems exhibit spin-glass ordering at low temperature, with static magnetic correlations at an IC wave vector,\cite{Tsunoda_92,Lamelas_95} and an up turn in resistivity at low temperature termed the Kondo effect.\cite{Mydosh_93}  The magnetic correlations are attributed to the RKKY interaction ({i.e.}, the coupling of local moments by conduction electrons), and the IC wave vector appears to correspond to $2k_F$, where $k_F$ is the Fermi wave vector.\cite{Lamelas_95}  

The present sample is located in the overdoped region, having a hole concentration of $p \sim 0.23$/Cu estimated from ARPES measurements.~\cite{Sato_unpub}  An ARPES study of (Bi,Pb)$_{2}$(Sr,La)$_{2}$CuO$_{6+y}$ by Kondo {\it et al.}\cite{Kondo_09} indicates that correlation effects, such as the pseudogap, are greatly reduced for overdoped samples.  That observation is consistent with neutron-scattering studies of LSCO indicating that antiferromagnetic correlations become quite weak with overdoping.\cite{waki_07,lips_07}  Thus, it is reasonable to consider the Fe spins doped into the metallic background of the overdoped BPSCO as Kondo scatterers that reduce the mobility of the charge carriers.\cite{Alloul_09,Balatsky_06}
The resistivity at low temperature varies as $\ln(1/T)$, which is consistent with Kondo behavior.
%
Scattering of the conduction electrons by the Fe moments could induce magnetic correlations at $2k_F$,
resulting in magnetic clusters. 
Compensation of the Fe moments by conduction electrons is, at best, only partial.  Magnetic fields would easily couple to the Fe moments, and suppress the $s-d$ exchange interaction between the local moments and conduction electrons, which is the source of both the Kondo effect and the RKKY interaction. %
Therefore the IC magnetic correlation is depressed by moderate magnetic fields, and the system shows the negative magnetoresistive effect.  The negative magnetoresistive effect is also discussed on the basis of the Kondo effect in Refs.~\onlinecite{Sekitani_03} and \onlinecite{ZnBi2201}.

\subsection{Stripe order}

There are several reports of impurity induced magnetic order in the under- and optimally-doped 214 compounds.~\cite{zn1,zn2,ni1,ni2}  In the latter cases, the induced order is a static stripe order around the impurity atoms owing to trapping of hole carriers by impurities, and the ordered regions appear to behave as  non-superconducting islands.~\cite{Nachumi_98}  We next consider whether this picture might be applicable to the present case.

Let us start with the large spin incommensurability of $\delta \sim 0.21$ found in Fe-doped BPSCO.  If we try to interpret this in terms of coupled charge and spin stripes, then the corresponding spin period of 5 Cu-O-Cu period would imply a charge stripe period of 2.5 lattice spacings.  That would require the charge and spin stripes to be not much wider than a single row of Cu atoms. 
Given that stripe order involves competition between strongly-correlated antiferromagnetism and the kinetic energy of the doped holes, it seems unlikely that such narrow domains
could be energetically favorable, especially in the overdoped regime.  
For reference,
the spin modulations in LSCO and Nd-doped LSCO tend to saturate at about 8 lattice spacings for $x>\frac18$.\cite{Birgeneau_JPSJ06}

The response to an applied magnetic field is another area of contrasting behavior.  Underdoped LSCO, La$_{2-x}$Ba$_{x}$CuO$_{4}$ (LBCO), and Nd-doped LSCO, particularly in the vicinity of 1/8 hole concentration, have a strong tendency towards static stripe order.  
Previous neutron scattering studies of these materials under magnetic fields revealed that the stripe order is enhanced for LSCO~\cite{Katano_00,Lake_02,Boris_02,Chang_08}, or stays constant for LBCO~\cite{Dunsiger_08,Wen_08} and Nd-doped LSCO~\cite{waki_03}; in any case, the stripe order is never weakened by fields up to at least 7~T.
These facts suggest that stripe order is robust against magnetic field, which is opposite to the present result.

Negative magnetoresistance has been observed in a limited number of cuprates, generally in cases where superconductivity is absent.  Examples include  lightly-doped LSCO,~\cite{Preyer_91} electron-doped thin films,~\cite{Sekitani_03} and Zn-doped Bi$_{2}$Sr$_{2-x}$La$_{x}$CuO$_{6+y}$~\cite{ZnBi2201}; in the case of La$_{1.79}$Eu$_{0.2}$Sr$_{0.01}$CuO$_4$, the decrease in $c$-axis resistivity due to an applied field was shown unambiguously to be due to spin scattering effects.\cite{Hucker_09}  More commonly, positive magnetoresistance is observed.\cite{Kimura_96,Balakirev_98} The salient point here is that negative magnetoresistance has not been observed in stripe ordered systems.  In the cases of LSCO, LBCO and Nd-doped LSCO with stripe order near the 1/8 hole concentration, the resistivity at low temperatures is generally increased due to the reduction in $T_c$ by applying a magnetic field, while the ``normal'' state resistivity above the zero-field $T_c$ is relatively insensitive to applied fields.\cite{Katano_00,Boe_96,Adachi_05,Li_07}  Thus, the negative magnetoresistance observed in Fe-doped BPSCO is another argument against a stripe interpretation.

\subsection{Remarks}

Based on the above discussion, we conclude that the present sample Bi$_{1.75}$Pb$_{0.35}$Sr$_{1.90}$Cu$_{0.91}$Fe$_{0.09}$O$_{6+y}$ is analogous to the dilute magnetic alloys, in which the Kondo behavior is relevant, and therefore the SR IC magnetic correlation induced by Fe-doping in the present ``overdoped'' BPSCO system originates from the correlation via polarization of conduction electron, the RKKY interaction.  The magnetic incommensurability should reflect the Fermi surface topology.  In contrast, magnetic or non-magnetic impurities doped in the ``underdoped'' cuprates induce stripe ordered state.  In that case, the incommensurability corresponds to the inverse of the stripe modulation period.  This difference between the underdoped and overdoped regimes might come from the difference in electronic nature: the former is in the strongly-correlated regime while the latter is in the metallic regime.

It is remarkable that the magnetic correlation in stripes in the underdoped superconducting region and that by the RKKY interaction in the present overdoped compound have the same modulation direction along the Cu-O-Cu bond.  Furthermore, the incommensurabilities for 
both cases 
follow
the $p=\delta$ relation.  
This raises the possibility of an interesting test of the nature of the magnetic excitations.  In the under- to optimally-doped regimes, the magnetic spectrum is characterized by an ``hour-glass'' dispersion.  There has been controversy over the the extent to which this spectrum derives from local moments or conduction electrons.  Can such a spectrum be observed in the present Fe-doped BPSCO sample?  If so, it would lend support to the argument that the conduction electrons are largely responsible for the magnetic response near optimal doping.  If not, it would be consistent with the view that hour-glass spectrum is a consequence of strongly-correlated antiferromagnetism.  Either way, measurements of the spin dynamics of this sample would yield interesting results.
%
%

\section{Summary}

Fe impurities in the overdoped BPSCO system induce SR IC magnetic correlation with unexpectedly large incommensurability $\delta=0.21$ below 40~K.  We have studied the magnetic field dependence of the magnetic correlation by neutron diffraction and the field dependence of the in-plane resistivity.  The magnetic peaks observed by neutron scattering show a small reduction in an applied magnetic field, and the resistivity shows a clear negative magnetoresistive effect below 40~K.  Such behavior is different from that typical of stripe ordered LSCO, LBCO and Nd-doped LSCO where the stripe order is robust against the magnetic fields.
The present results show greater similarities to dilute magnetic alloys, in which the Kondo effect is relevant.  The Fe spins in the overdoped metallic background produces a Kondo effect which results in the up turn in the resistivity at low temperature in the form of $\ln(1/T)$.  On the other hand, SR magnetic correlation is induced by the RKKY interaction.  
Magnetic fields couple to the Fe moments, competing with the exchange interaction between the local moments and conduction electrons.  This effect disturbs both the Kondo effect and the RKKY interaction, resulting in the reduction of the spatially-modulated magnetic correlations and the negative magnetoresistive effect.

\begin{acknowledgments}

We thank K. Kaneko, M. Matsuda, M. Fujita, and J. A. Fernandez-Baca for invaluable discussion.  We also thank S. Okayasu for his help in SQUID measurements.
This work is part of the US-Japan Cooperative Program on Neutron scattering.  The work at the HFIR at ORNL was partially funded by the Division of Scientific User Facilities, Office of Basic Energy Science, US Department of Energy.  The study performed at JRR-3 at Tokai was carried out under the Common-Use Facility Program of JAEA, and the Quantum Beam Technology Program of JST.  Magnetoresistive measurement was performed at the High Field Laboratory for Superconducting Materials, Institute for Materials Research, Tohoku University.  We acknowledge financial support by Grant-in-Aid from the Ministry of Education, Culture, Sports, Science and Technology.  JMT is supported by the U.S. Department of Energy, Office of Basic Energy Sciences, Division of Materials Sciences and Engineering, under Contract No.~DE-AC02-98CH110886.

\end{acknowledgments}


\begin{thebibliography}{99}

\bibitem{Birgeneau_JPSJ06}
R. J. Birgeneau, C. Stock, J. M. Tranquada, and K. Yamada,
J. Phys. Soc. Jpn. {\bf 75}, 111003 (2006).

\bibitem{hayd04}
S.~M. Hayden, H.~A. Mook, P. Dai,  T.~G. Perring, and F. Do\u{g}an,
Nature {\bf 429}, 531 (2004).

\bibitem{tran04}
J.~M. Tranquada, H. Woo, T.~G. Perring, H. Goka, G.~D. Gu, G. Xu, M. Fujita, and K. Yamada,
Nature {\bf 429}, 534 (2004).

\bibitem{waki_04}
S. Wakimoto, H. Zhang, K. Yamada, I. Swainson, H.-K. Kim, and R.J. Birgeneau,
Phys. Rev. Lett. {\bf 92}, 217004 (2004).

\bibitem{waki_07}
S. Wakimoto, K. Yamada, J. M. Tranquada, C. D. Frost, R. J. Birgeneau, and H. Zhang,
Phys. Rev. Lett. {\bf 98}, 247003 (2007).

\bibitem{lips_07}
O. J. Lipscombe, S. M. Hayden, B. Vignolle, D. F. McMorrow, and T. G. Perring,
Phys. Rev. Lett. {\bf 99}, 067002 (2007).

\bibitem{xu_cm09}
Guangyong Xu, G. D. Gu, M. Hucker, B. Fauque, T. G. Perring, L. P. Regnault, and J. M. Tranquada,
Nature Phys. {\bf 5}, 642 (2009).

\bibitem{Bi}
H. F. Fong, P. Bourges, Y. Sidis, L. P. Regnault, A. Ivanov, G. D. Gu, N. Koshizuka and B. Keimer,
Nature {\bf 398}, 588 (1999).

\bibitem{Tl}
H. He, P. Bourges, Y. Sidis, C. Ulrich, L. P. Regnault, S. Pailhes, N. S. Berzigiarova, N. N. Kolesnikov, and B. Keimer,
Science {\bf 295}, 1045 (2002).

\bibitem{Fujita_09CM}
M. Fujita, M. Enoki, S. Iikubo, K. Kudo, N. Kobayashi, and K. Yamada,
arXiv:0903.5391[e-print arXiv].

\bibitem{hira_09}
H. Hiraka, Y. Hayashi, S. Wakimoto, M. Takeda, K. Kakurai, T. Adachi, 
Y. Koike, I. Yamada, M. Miyazaki, M. Hiraishi, S. Takeshita, A. Kohda, R. Kadono, J. M. Tranquada, and K. Yamada,
Phys. Rev. B {\bf 81}, 144501 (2010).

\bibitem{Yamada_98}
K. Yamada, C. H. Lee, K. Kurahashi, J. Wada, S. Wakimoto, S. Ueki, H. Kimura, Y. Endoh, S. Hosoya, G. Shirane, R. J. Birgeneau, M. Greven, M. A. Kastner, and Y. J. Kim,
Phys. Rev. B {\bf 57}, 6165 (1998).


\bibitem{zn1}
H. Kimura, K. Hirota, H. Matsushita, K. Yamada, Y. Endoh, S.-H. Lee, C. F. Majkrzak, R. Erwin, G. Shirane, M. Greven, Y. S. Lee, M. A. Kastner, and R. J. Birgeneau,
Phys. Rev. B {\bf 59}, 6517 (1999).

\bibitem{zn2}
I. Watanabe, T. Adachi, K. Takahashi, S. Yairi, Y. Koike, and K. Nagamine,
Phys. Rev. B {\bf 65}, 180516(R) (2002).

\bibitem{ni1}
M. Kofu, H. Kimura, and K. Hirota,
Phys. Rev. B {\bf 72}, 064502 (2005).

\bibitem{ni2}
T. Adachi, N. Oki, Risdiana, S. Yairi, Y. Koike, and I. Watanabe,
Phys. Rev. B {\bf 78}, 134515 (2008).

\bibitem{Katano_00}
S. Katano, M. Sato, K. Yamada, T. Suzuki, and T. Fukase,
Phys. Rev. B {\bf 62}, R14677 (2000).

\bibitem{Lake_02}
B. Lake, H. M. R\/onnow, N. B. Christensen, G. Aeppli, K. Lefmann, D. F. McMorrow, P. Vorderwisch, P. Smeibidl, N. Mangkorntong, T. Sasagawa, M. Nohara, T. Takagi, and T. E. Mason,
Nature {\bf 415}, 299 (2002).

\bibitem{Boris_02}
B. Khaykovich, Y. S. Lee, R. Erwin, S.-H. Lee, S. Wakimoto, K. J. Thomas, M. A. Kastner, and R. J. Birgeneau,
Phys. Rev. B {\bf 66}, 014528 (2002).

\bibitem{Chang_08}
J. Chang, Ch. Niedermayer, R. Gilardi, N. B. Christensen, H. M. Ronnow, D. F. McMorrow, M. Ay, J. Stahn, O. Sobolev, A. Hiess, S. Pailhes, C. Baines, N. Momono, M. Oda, M. Ido, and J. Mesot,
Phys. Rev. B {\bf 78}, 104525 (2008).


\bibitem{Sato_unpub}
T. Sato, private communications.

\bibitem{Tsunoda_92}
Y. Tsunoda and J. W. Cable,
Phys. Rev. B {\bf 46}, 930 (1992).

\bibitem{Lamelas_95}
F. J. Lamelas, S. A. Werner, S. M. Shapiro, and J. A. Mydosh,
Phys. Rev. B {\bf 51}, 621 (1995).

\bibitem{Mydosh_93}
J. A. Mydosh,
{\it Spin Glasses: An Experimental Introduction} (Taylor \&\ Francis, London, 1993).

\bibitem{Kondo_09}
T. Kondo, R. Khasanov, T. Takeuchi, J. Schmalian, and A. Kaminski,
Nature {\bf 457}, 296 (2009).

\bibitem{Alloul_09}
H. Alloul, J. Bobroff, M. Gabay, and P. J. Hirschfeld,
Rev. Mod. Phys. {\bf 81}, 45 (2009).

\bibitem{Balatsky_06}
A. V. Balatsky, I. Vekhter, and J.-X. Zhu,
Rev. Mod. Phys. {\bf 78}, 373 (2006).


\bibitem{Nachumi_98}
B. Nachumi, Y. Fudamoto, A. Keren, K. M. Kojima, M. Larkin, G. M. Luke, J. Merrin, O. Tchernyshyov, Y. J. Uemura, N. Ichikawa, M. Goto, H. Takagi, S. Uchida, M. K. Crawford, E. M. McCarron, D. E. MacLaughlin, and R. H. Heffner,
Phys. Rev. B {\bf 58}, 8760 (1998).

\bibitem{Dunsiger_08}
S. R. Dunsiger, Y. Zhao, Z. Yamani, W. J. L. Buyers, H. A. Dabkowska, and B. D. Gaulin,
Phys. Rev. B {\bf 77}, 224410 (2008).

\bibitem{Wen_08} J. Wen, Z. Xu, G. Xu, J. M. Tranquada, G. Gu, S. Chang and H. J. Kang,
Phys. Rev. B {\bf 78}, 212506 (2008).

\bibitem{waki_03}
S. Wakimoto, R. J. Birgeneau, Y. Fujimaki, N. Ichikawa, T. Kasuga, Y. J. Kim, K. M. Kojima, S.-H. Lee, H. Niko, J. M. Tranquada, S. Uchida, and M. v. Zimmermann,
Phys. Rev. B {\bf 67}, 184419 (2003).

\bibitem{Preyer_91}
N. W. Preyer, M. A. Kastner, C. Y. Chen, R. J. Birgeneau, and Y. Hidaka,
Phys. Rev. B {\bf 44}, 407 (1991).

\bibitem{Sekitani_03}
Tsuyoshi Sekitani, Michio Naito, and Noboru Miura,
Phys. Rev. B {\bf 67}, 174503 (2003).

\bibitem{ZnBi2201}
Y. Hanaki, Y. Ando, S. Ono, and J. Takeya,
Phys. Rev. B {\bf 64}, 172514 (2001).

\bibitem{Hucker_09}
M. H\"ucker,
Phys. Rev. B {\bf 79}, 104523 (2009).

\bibitem{Kimura_96}
T. Kimura, S. Miyasaka, H. Takagi, K. Tamasaku, H. Eisaki, S. Uchida, K. Kitazawa, M. Hiroi, M. Sera, and N. Kobayashi,
Phys. Rev. B {\bf 53}, 8733 (1996).

\bibitem{Balakirev_98}
F. F. Balakirev, I. E. Trofimov, S. Guha, M. Z. Cieplak, and P. Lindenfeld,
Phys. Rev. B {\bf 57}, R8083 (1998).

\bibitem{Boe_96}
G. S. Boebinger, Y. Ando, A. Passner, T. Kimura, M. Okuya, J. Shimoyama, K. Kishio, K. Tamasaku, N. Ichikawa, and S. Uchida,
Phys. Rev. Lett. {\bf 77}, 5417 (1996).

\bibitem{Adachi_05}
T. Adachi, N. Kitajima, T. Manabe, Y. Koike, K. Kudo, T. Sasaki, and N. Kobayashi,
Phys. Rev. B {\bf 71}, 104516 (2005).

\bibitem{Li_07}
Q. Li, M. H\"ucker, G. D. Gu, A. M. Tsvelik, and J. M. Tranquada,
Phys. Rev. Lett. {\bf 99}, 067001 (2007).



\end{thebibliography}
\end{document}